\documentclass[aps,pre,onecolumn,showpacs,superscriptaddress]{revtex4-2}

\usepackage{amsmath,amssymb,amsfonts}
\usepackage{graphicx}
\usepackage{bm}
\usepackage{hyperref}
\usepackage{color}
\usepackage{mathrsfs}

\begin{document}

\title{Random planting with harvest: A statistical--mechanical analysis}

\author{Julian Talbot}
\affiliation{Sorbonne Universit\'e, CNRS, Laboratoire de Physique Th\'eorique de la Mati\`ere Condens\'ee (UMR CNRS 7600), 4 Place Jussieu, 75005 Paris, France}

\date{\today}

\begin{abstract}

We formulate a statistical–mechanical description of a recently introduced random planting model in which plants are represented by growing hard disks. Seedlings of negligible size are introduced at random positions in a field, grow at a prescribed rate, and are harvested upon reaching a fixed maturity diameter. Planting attempts that would lead to an overlap at any time during growth are rejected. Starting from an empty field, this simple dynamical rule drives the system to a nonequilibrium steady state in which the mean planting and harvesting rates coincide. We show that the steady state can be mapped onto a nonadditive polydisperse hard–disk fluid and exploit this mapping to develop analytical predictions based on a low-density virial expansion and on scaled particle theory. The resulting description yields an effective adsorption isotherm for the steady-state plant density as a function of the planting rate and compares favorably with numerical simulations over a wide range of parameters. At large planting rates, the density approaches the optimal value achieved by desynchronized regular planting, and the data are consistent with an algebraic approach to this limit with an exponent close to $1/3$. Beyond density and yield, we show that the spatial organization of the field at high planting rates exhibits clear signatures of the same underlying geometric constraints that characterize optimal desynchronized planting. This connection is revealed through both the conventional radial distribution function and a radius–resolved pair correlation $g(z,r)$ which highlights strong size correlations associated with parent–child seeding events and whose structure can be interpreted as a dynamically broadened precursor of the corresponding ideal mixed-size lattice. Finally, we extend the theory to sigmoidal growth laws and compute the associated virial coefficient.
\end{abstract}

\maketitle

\section{Introduction}
\label{sec:intro}

Agroecology has emerged as a framework for designing sustainable food ecosystems that reduce the environmental costs of industrial agriculture~\cite{Wezel2009agroecology}. A recurring challenge in this setting is to understand how simple geometric constraints—finite plant size, growth, and local competition for space—translate into limits on achievable yield. The random planting model (RPM) introduced by Colliaux \emph{et al.}~\cite{Colliaux2023models} provides a minimal and analytically tractable starting point: plants are idealized as hard disks that grow deterministically from seedlings to a fixed harvest diameter and are removed after a lifetime $\tau$. Planting attempts occur at random positions but are rejected whenever they would lead to an overlap at any time in the future. Starting from an empty field, this rule drives the system to a nonequilibrium steady state in which the mean planting and harvesting rates coincide. Despite its simplicity, the RPM generates nontrivial spatial organization, making it a useful setting for isolating how growth, exclusion, and harvesting jointly determine yield. Related questions for real crops have been explored using empirical and mechanistic approaches in Refs.~\cite{Deng2012,deng2012insights}. From an agronomic perspective, the central observable is the yield, defined as the number of plants harvested per unit time and unit area as a function of the planting rate. Because each plant lives for a fixed time $\tau$, the steady-state yield is proportional to the steady-state density, so bounding or predicting the density directly constrains yield. One is also interested in the spatial organization of the field and, for duocultures, the relative fraction of harvested species. While real growth is typically sigmoidal, it is natural to begin with linear growth, for which many calculations can be carried out explicitly.

A useful benchmark for the RPM is regular planting, where seedlings are sown at prescribed spacings. In synchronized planting, all seedlings are introduced simultaneously and harvested simultaneously one lifetime later. Higher yield can be achieved by desynchronized planting, in which a second layer is introduced after a delay ($\tau/2$ maximizes the yield) on interstitial sites opened by the growth schedule. The large planting rate behavior of the RPM can be compared against such optimized desynchronized protocols, which provide a natural geometric upper bound.

\begin{figure}[t]
 \centering
 \includegraphics[width=0.95\linewidth]{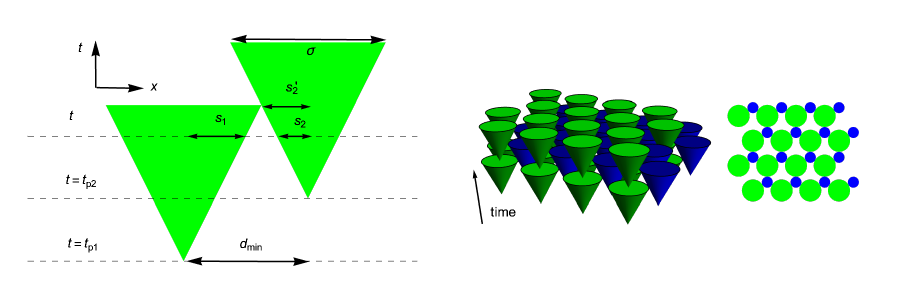}
 \caption{Left: Spacetime diagram of the random planting model. The older plant (left) was planted at $t_{p1}$ and will be harvested at $t=t_{p1}+\tau$. In order to avoid overlap the younger plant must be placed at a distance of at least $d_{\rm min}=\sigma-| s_2-s_1|$.  Middle: Optimum planting configuration for linear growth that is obtained for a desynchronisation factor of $\Delta t/\tau=1/2$.  The density is $\rho_{\rm max}\sigma^{2}=64/(27\sqrt{3})\approx 1.3685$. The spacetime representation shows three consecutive layers; Right: overhead view of the field at time $t=\tau/2$. The plants in the first layer (green)  have reached maturity and are about to be harvested, while the second layer plants (blue) are half way to maturity ($s=\sigma/4$).}
 \label{fig:cones}
\end{figure}

A one--dimensional version of the RPM, in which plants grow on a line of fixed length, was studied in Ref.~\cite{talbot2025random} using an event--driven algorithm. The advantage of this algorithm is that it avoids rejected insertion attempts, thereby allowing high effective planting rates to be reached at modest computational cost. For a single species, the steady--state yield approaches a maximum of $4/3$ plants per unit length and unit time, a value that can be explained by elementary geometric arguments~\cite{talbot2025optimising}. Remarkably, the field structure in this limit corresponds to a regular desynchronized planting. In two dimensions, for plants that have diameter $\sigma$ at harvest, the optimal density for linear growth is~\cite{talbot2025optimising}

\begin{equation}
 \rho_{\max}\sigma^2 = \frac{64}{27\sqrt{3}} \simeq 1.3685\,,
\end{equation}
achieved by desynchronized periodic planting: an initial batch is placed on a hexagonal lattice and, after a time $\tau/2$, a second batch is inserted on the interstitial sites (See Fig. \ref{fig:cones}).

Beyond its possible application to agroecology, the RPM is interesting from the viewpoint of nonequilibrium statistical mechanics. It is closely related to several classical models involving packing processes, in particular the equilibrium hard--disk fluid~\cite{hansen2013theory}, random sequential adsorption (RSA) and irreversible deposition models~\cite{evans1993random,schaaf2000random,Talbot2000,wang1994fast,penrose2002limit,Kubala2022}, adsorption--desorption (parking--lot) models~\cite{Talbot2000b,Oleyar2007}, and packing--limited growth (PLG) models~\cite{Andrienko1994,Dodds2002,dodds2003packing}. The present work exploits these connections to develop a quantitative description of the steady state of the two--dimensional RPM.

The main ideas are as follows. First, we show that the exclusion rule of the RPM can be cast as an effective pair potential between polydisperse disks, which is nonadditive because the interaction range depends on the age difference between plants. Second, we relate the steady--state planting probability to the excess chemical potential of a tagged seedling, thereby mapping the steady state onto an adsorption isotherm of a polydisperse hard--disk fluid. At low densities this mapping may be treated using a virial expansion. For higher densities we adapt scaled particle theory (SPT)~\cite{reiss1959statistical,helfand1961theory,lebowitz1965scaled} to construct a semi--empirical equation of state with a single effective diameter parameter that encapsulates polydispersity and nonadditivity. We show that this description yields accurate predictions of the steady--state density over a wide range of planting rates. At very large planting rates the density approaches $\rho_{\max}$, and the simulations are consistent with an algebraic approach $\rho_{\max}-\rho \propto k_p^{-b}$ with $b\simeq 1/3$.

We also characterize the structure of the field. Besides the usual radial distribution function $g(r)$, we introduce a radius--resolved pair correlation $g(z,r)$, where $z$ is the absolute difference in the radii of two plants. This quantity reveals strong size correlations associated with the seeding of new plants in the shrinking exclusion halos of older ones, leading to a prominent peak near $z \simeq \sigma /4$ at short separations when the planting rate is large.

Finally, we relax the assumption of linear growth and consider a logistic growth law. We determine the steady--state size distribution and compute the corresponding average second virial coefficient, which can be used to obtain low--density predictions for the steady--state density.

The paper is organized as follows. In Sec.~\ref{sec:model} we define the RPM and the simulation procedure. Section~\ref{sec:transient} describes the transient dynamics from an initially empty field. Section~\ref{sec:steady} develops the analytical description of the steady state, based on a mean--field kinetic equation, a low--density virial expansion, and an SPT--based approximation valid up to moderate densities. Section~\ref{sec:structure} discusses the spatial correlations in the steady state. Section~\ref{sec:sigmoidal} extends the theory to sigmoidal growth. We conclude in Sec.~\ref{sec:discussion} with a summary and some open questions.

\section{Model and simulation}
\label{sec:model}

We consider a square field of area $L^2$ with periodic boundary conditions, in which plants are represented by circular disks. A plant born at time $t_0$ has a radius $s(t)$ that increases deterministically with time. In most of the paper we assume linear growth,
\begin{equation}
 s(t) = \frac{\sigma}{2}\frac{t-t_0}{\tau}\,,
 \qquad 0 \le t-t_0 \le \tau,
\end{equation}
so that the disk diameter reaches $\sigma$ at the harvest time $t_0+\tau$, at which point the plant is removed from the system. We measure lengths in units of $\sigma$ and times in units of $\tau$ when convenient.

Planting attempts occur at a fixed time interval $\Delta t$. At each attempt a position $\bm{r}$ is drawn from a random, uniform distribution in the simulation cell. The attempt is accepted if and only if the trajectory of the seedling does not lead to an overlap with any existing plant at any time between its birth and its harvest. Since all plants grow at the same rate, this requirement can be formulated purely in terms of their radii at the time of insertion.

Let $\bm{r}_1$ and $\bm{r}_2$ denote the positions of the centers of two plants with radii $s_1$ and $s_2$ at the instant of the attempt, and let $d_{12} = \lVert \bm{r}_1-\bm{r}_2\rVert$ be their separation (minimum--image convention). Because their radii subsequently increase at the same rate, their difference $|s_1-s_2|$ remains constant until one of them is harvested. The sum $s_1+s_2$ is maximal when the oldest reaches maturity. From Fig. (\ref{fig:cones})  one sees that  $d_{\rm min}=\sigma/2+s_2'$, where $s_2'$ is the radius of the younger plant at the instant that the older one is harvested. The condition that the two plants never overlap during their coexistence is therefore
\begin{equation}
 d_{12} > d_{\rm min}=\sigma - |s_1-s_2|\,.
 \label{eq:hard-core}
\end{equation}
The minimum allowed distance is $\sigma/2$, which occurs for a seedling ($s_1=0$) and a mature plant ($s_2=\sigma/2$), while the maximum exclusion distance $\sigma$ arises when the two plants are of the same age and hence have identical radii.

If Eq.~\eqref{eq:hard-core} is satisfied with respect to all existing plants in the field, the attempt is accepted and a new seedling with $s=0$ and position $\bm{r}$ is created. Otherwise the trial is rejected. After each planting attempt the time is advanced by $\Delta t$, the radii of all existing plants are updated, and plants that have reached the maturity radius $\sigma/2$ are removed. The planting rate for an empty field is thus
\begin{equation}
 k_p = \frac{1}{L^2 \Delta t}\,,
 \label{eq:kp-def}
\end{equation}
measured in units of attempts per unit area per unit time. We focus on the single--species case.

\begin{figure}[t]
 \centering
 \includegraphics[width=0.95\linewidth]{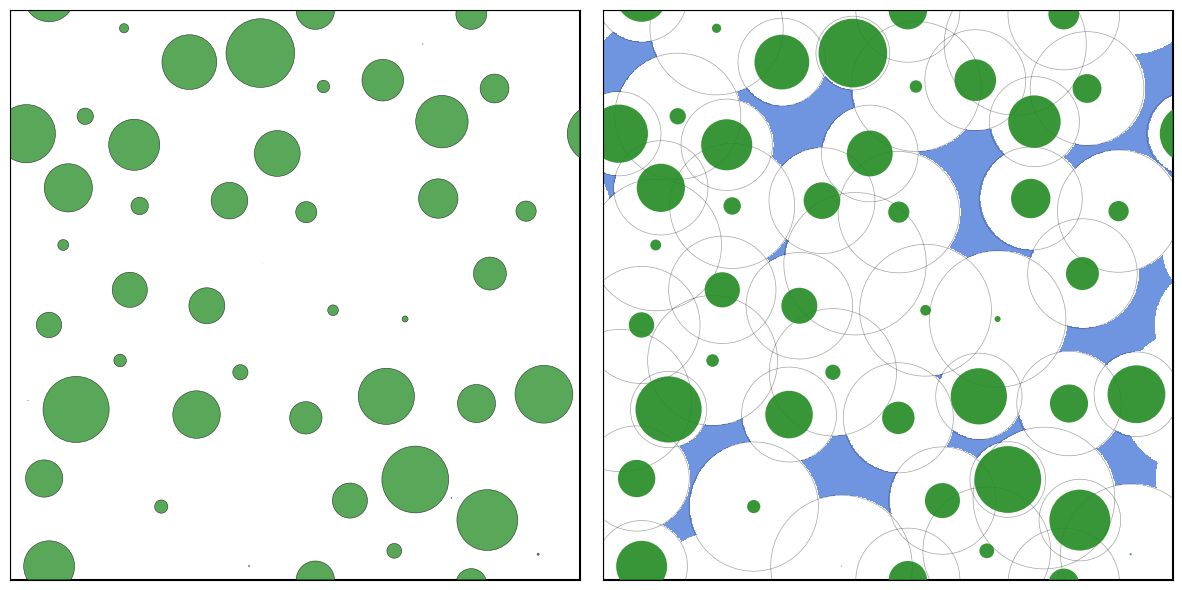}
 \caption{Configuration of plants in the steady state for $L/\sigma = 8$ and $k_p=10$. 
 Left: plant disks only. Right: same configuration showing the exclusion circles (light gray) associated with each plant and the instantaneous available surface (blue) into which a new seedling can be inserted without causing future overlaps.}
 \label{fig:configs}
\end{figure}

Starting from an empty field, the density of plants $\rho(t)$ increases and, after a transient, reaches a steady state in which the mean rate of planting equals the mean rate of harvest. Figure~\ref{fig:configs} shows a typical configuration in the steady state and the corresponding exclusion regions.

In the steady state, plant ages and hence radii are continuously distributed. For linear growth the lifetime $\tau$ is fixed and the birth process is stationary, so the probability density function of radii, $\psi(s)$, is uniform between $0$ and the maturity radius $\sigma/2$:
\begin{equation}
 \psi(s) = \frac{2}{\sigma}, \qquad 0 \le s \le \frac{\sigma}{2}.
 \label{eq:psi-linear}
\end{equation}

\section{Transient dynamics}
\label{sec:transient}

\begin{figure}[t]
 \centering
 \includegraphics[width=0.9\linewidth]{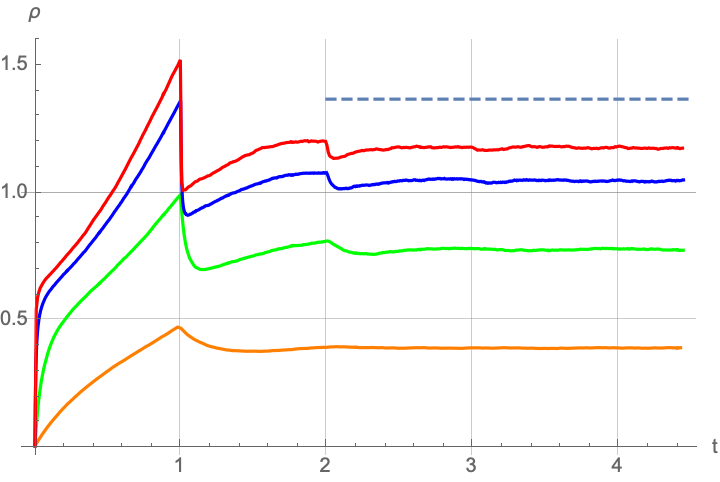}
 \caption{Transient evolution of the plant density $\rho(t)$ from an initially empty field for planting rates $k_p=1,10,100,500$ (bottom to top). Time is measured in units of the plant lifetime~$\tau$. The dashed horizontal line marks the theoretical upper bound $\rho_{\max}=64/(27\sqrt{3})$.}
 \label{fig:transients}
\end{figure}

The early--time evolution of the density from an initially empty field depends strongly on the planting rate $k_p$. Figure~\ref{fig:transients} shows $\rho(t)$ for several values of $k_p$ obtained from simulations with $L/\sigma=15$, each curve representing an average over $50$--$100$ independent realizations.

For moderate planting rates ($k_p\sim 1$) the density increases sublinearly and reaches a maximum at $t\simeq \tau$. This maximum is followed by a damped oscillation before the steady state is attained. The oscillation reflects the finite lifetime of plants: a first cohort of seedlings is planted between $t=0$ and $t\simeq\tau$, and its harvest at $t\simeq\tau$ produces a noticeable drop in the density.

For very large planting rates the dynamics is qualitatively different. The density initially rises rapidly as many seedlings are accepted, until the field becomes crowded and almost all new attempts are rejected. At this stage the configuration of exclusion regions is close to that of a jamming configuration in an RSA process of hard disks with diameter $\sigma$~\cite{evans1993random,schaaf2000random,Talbot2000,wang1994fast,Kubala2022}. The corresponding coverage is $\theta_{\rm RSA}\simeq 0.547$ and the associated density is
\begin{equation}
 \rho_{\rm RSA} = \frac{4\theta_{\rm RSA}}{\pi\sigma^2} \simeq 0.696\,\sigma^{-2}.
\end{equation}
In the simulation for $k_p=500$ a shoulder appears in $\rho(t)$ around $\rho\simeq 0.6$, close to this value. At that point the available area for new seedlings has essentially vanished. However, as time progresses the plants continue to grow, their exclusion radii shrink, and previously blocked regions become accessible. The density thus increases further until the first plants reach maturity at $t=\tau$, whereupon the simultaneous harvest of the earliest cohort produces a sharp drop in $\rho(t)$, reminiscent of features observed in deposition models with coupled nucleation and growth~\cite{khanam2019effects}.

\begin{figure}[t]
 \centering
 \includegraphics[width=0.9\linewidth]{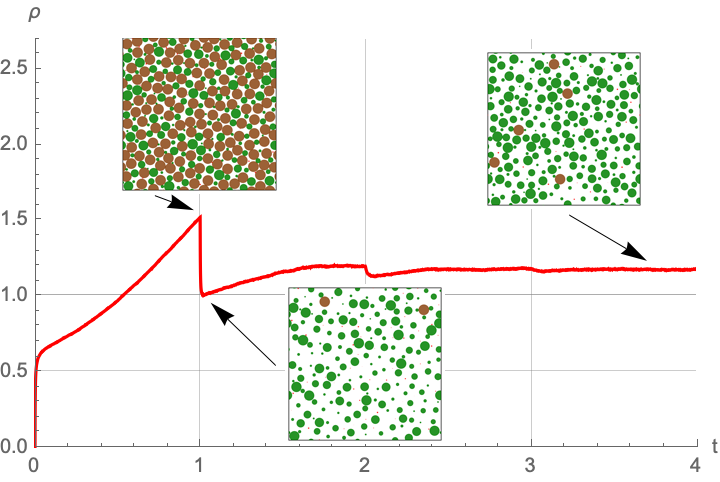}
 \caption{Snapshots of the field morphology for $k_p=500$ at different times, illustrating the transient dynamics. Colors distinguish young plants (red), intermediate ages (green) and nearly mature plants (brown).}
 \label{fig:kp500_snapshots}
\end{figure}

Figure~\ref{fig:kp500_snapshots} illustrates the morphology of the field at different times for a large planting rate. Shortly before $t=\tau$ the field resembles an RSA jamming configuration with a broad distribution of radii. Just after $t=\tau$ many of the oldest plants are harvested and the field becomes more open. At later times the system approaches the steady state characterized in Sec.~\ref{sec:steady}.

\section{Steady--state theory}
\label{sec:steady}

\subsection{Mean--field kinetic equation}

In the steady state the density is time independent on average and the mean planting rate equals the mean harvesting rate. Let $\phi(\rho)$ denote the probability that a planting attempt is accepted given an instantaneous density~$\rho$. The effective planting rate is then $k_p\phi(\rho)$. Because each plant lives for a time~$\tau$, the mean harvesting rate is $\rho/\tau$. The density obeys the mean--field kinetic equation
\begin{equation}
 \frac{{\rm d}\rho}{{\rm d}t} = k_p \phi(\rho) - \frac{\rho}{\tau}.
 \label{eq:kinetic}
\end{equation}
In the stationary regime ${\rm d}\rho/{\rm d}t=0$ and
\begin{equation}
 k_p \tau \phi(\rho) = \rho.
 \label{eq:isotherm}
\end{equation}
This equation is formally analogous to an adsorption isotherm, with $k_p\tau$ playing the role of a reduced chemical potential and $\phi$ that of an insertion probability.

Our task is therefore to determine the insertion probability $\phi(\rho)$ for a seedling in terms of the instantaneous density~$\rho$. The key observation is that Eq.~\eqref{eq:hard-core} can be interpreted as defining an effective pair potential $u(r,s_1,s_2)$ between disks of radii $s_1$ and $s_2$:
\begin{equation}
u(r,s_1,s_2) =
\begin{cases}
\infty, & r < \sigma - |s_1-s_2|,\\
0,      & r \ge \sigma - |s_1-s_2|,
\end{cases}
\label{eq:pair-pot}
\end{equation}
where $r$ is the center--to--center separation. The interaction range is nonadditive because it depends on the difference of radii rather than their sum. The seedling may be treated as a tagged species with radius $s=0$, immersed in a polydisperse hard--disk fluid with radius distribution $\psi(s)$.

\subsection{Low--density virial expansion}

At low density the excess chemical potential $\mu^{\rm ex}$ of the tagged seedling can be expressed as a virial series in the density~$\rho$,
\begin{equation}
 \beta \mu^{\rm ex} = 2 B_2 \rho + \frac{3}{2} B_3 \rho^2 + \cdots,
 \label{eq:mu-vir}
\end{equation}
where $\beta = 1/(k_{\rm B}T)$ and $B_k$ are the virial coefficients. The insertion probability is related to the excess chemical potential by
\begin{equation}
 \phi = \exp(-\beta \mu^{\rm ex}).
 \label{eq:phi-mu}
\end{equation}
Expanding Eq.~\eqref{eq:phi-mu} to first order in $\rho$ yields
\begin{equation}
 \phi(\rho) = 1 - 2 B_2 \rho + O(\rho^2).
\end{equation}

For the present system the second virial coefficient $B_2$ is obtained by integrating the Mayer function associated with the pair potential~\eqref{eq:pair-pot} between a seedling ($s_1=0$) and a plant of radius~$s$,
\begin{equation}
 B_2(s) = -\frac{1}{2}\int_0^\infty \bigl( e^{-\beta u(r,s)}-1\bigr)\,2\pi r\,{\rm d} r 
        = \frac{\pi}{2} (\sigma - s)^2,
 \label{eq:B2s}
\end{equation}
where $u(r,s)\equiv u(r,0,s)$. Averaging over the radius distribution $\psi(s)$ yields the effective second virial coefficient
\begin{equation}
 B_2 = \int_0^{\sigma/2} B_2(s)\,\psi(s)\,{\rm d}s.
 \label{eq:B2-avg}
\end{equation}
For linear growth and the uniform distribution~\eqref{eq:psi-linear} this gives
\begin{equation}
 B_2 = \frac{7\pi}{24} \sigma^2.
 \label{eq:B2-linear}
\end{equation}

\begin{figure}[t]
 \centering
 \includegraphics[width=0.9\linewidth]{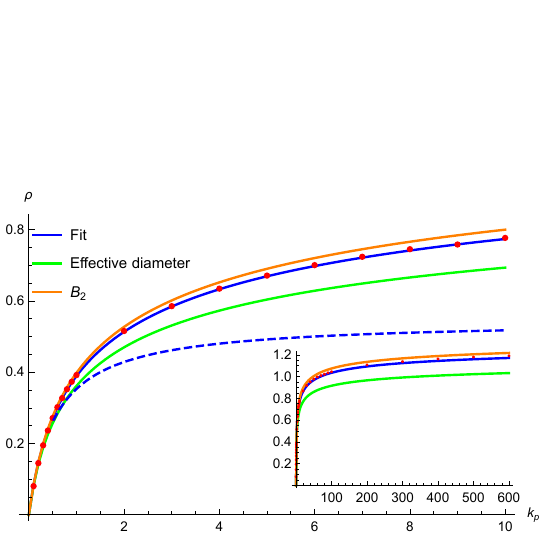}
 \caption{Steady--state density $\rho$ as a function of the planting rate $k_p$ for $L/\sigma=20$. Points: simulation data. Dashed line: low--density prediction~\eqref{eq:rho-low-density}. Solid lines: SPT--based models described in Sec.~\ref{subsec:SPT}.}
 \label{fig:yield}
\end{figure}

Combining Eq.~\eqref{eq:isotherm} with the low--density form of $\phi$ leads to the explicit expression
\begin{equation}
 \rho \simeq \frac{k_p\tau}{1 + 2 B_2 k_p\tau}
            = \frac{k_p\tau}{1 + \frac{7}{12}\pi \sigma^2 k_p\tau}.
 \label{eq:rho-low-density}
\end{equation}
As shown in Fig.~\ref{fig:yield}, this expression provides a good description of the steady--state density for $k_p\lesssim 1$, but it increasingly underestimates the density at larger planting rates, as is typical of low-order virial truncations.

\subsection{Scaled particle theory}
\label{subsec:SPT}

To extend the description to higher densities we adapt Scaled Particle Theory (SPT), originally developed by Reiss, Frisch and Lebowitz for hard spheres~\cite{reiss1959statistical} and generalized to hard disks and mixtures in Refs.~\cite{helfand1961theory,lebowitz1965scaled}. SPT and related approaches continue to be refined for mixtures and polydisperse systems~\cite{santos2001virial,santos2005equation,santos2017equation,hansen2019scaling} and have also been applied to multicomponent adsorption kinetics and polydisperse adsorption~\cite{talbot1994new,olson2000equilibria}. For a monodisperse hard--disk fluid of diameter~$\sigma$ at density~$\rho$, the SPT expression for the probability of inserting a test disk of radius $r$ is
\begin{equation}
 \phi_{\rm SPT}(\rho,r) = (1-\theta)\,
 \exp\!\left[
  - A(r)\rho + \frac{r\gamma}{1-\theta}
  - \frac{A(r)\gamma^2}{4\pi (1-\theta)^2}
 \right],
 \label{eq:SPT-general}
\end{equation}
where 
\begin{equation}
 \theta = \frac{\pi}{4}\rho \sigma^2, \qquad
 \gamma = \pi \rho \sigma
\end{equation}
are the coverage and total circumference of the fluid disks, and $A(r)=\pi r^2$ is the area of the test particle.

In the RPM the effective interaction~\eqref{eq:pair-pot} is nonadditive and the fluid is polydisperse, so Eq.~\eqref{eq:SPT-general} cannot be applied directly. Nonadditive and polydisperse hard--sphere and hard--disk mixtures have been studied extensively~\cite{santos2001virial,santos2005equation,santos2017equation,pellicane2022theory}, but implementing a fully general nonadditive mixture theory here would be unnecessarily involved for our purposes. We nonetheless expect that a monodisperse SPT form with an appropriately renormalized diameter can capture much of the behavior. Specifically, we evaluate Eq.~\eqref{eq:SPT-general} for a test disk of radius $\sigma/2$ and replace $\sigma$ in the definition of the coverage by an effective diameter $\sigma' = a\sigma$, while keeping the actual density~$\rho$. This yields
\begin{equation}
 \phi_{\rm SPT}(\rho) \equiv \phi_{\rm SPT}(\rho,\sigma/2) 
 = (1-\theta') \exp\!\left[
  -\frac{3\theta'}{1-\theta'} - \frac{\theta'^2}{(1-\theta')^2}
 \right],
 \label{eq:SPT-sigma-half}
\end{equation}
with
\begin{equation}
 \theta' = \frac{\pi}{4} \rho (a\sigma)^2.
\end{equation}
Substituting $\phi_{\rm SPT}$ into Eq.~\eqref{eq:isotherm} yields an implicit equation for $\rho$ at fixed $k_p\tau$, which we solve numerically.

We have considered three ways of choosing the dimensionless parameter~$a$:
\begin{enumerate}
 \item[(i)] \emph{Mean effective diameter.} We treat the fluid as monodisperse with an effective hard--core distance
 \begin{equation}
  \sigma_{\rm eff} = \sigma - \langle |s_1-s_2|\rangle,
 \end{equation}
 where the average is taken over two independent radii sampled from $\psi(s)$. For the uniform distribution~\eqref{eq:psi-linear} this yields $\sigma_{\rm eff} = (5/6)\sigma$, so $a=5/6\simeq 0.8333$.

 \item[(ii)] \emph{Matching the second virial coefficient.} At low density the insertion probability obtained from Eq.~\eqref{eq:SPT-sigma-half} has the expansion $\phi_{\rm SPT}(\rho)=1-\pi a^2 \rho + O(\rho^2)$. Matching this with the exact low--density result $1-2B_2\rho$, with $B_2$ given by Eq.~\eqref{eq:B2-linear}, gives
 \begin{equation}
  a = \sqrt{\frac{7}{12}} \simeq 0.7638.
 \end{equation}

 \item[(iii)] \emph{Fit to simulation data.} We treat $a$ as an adjustable parameter and fit Eq.~\eqref{eq:isotherm} with $\phi(\rho)=\phi_{\rm SPT}(\rho)$ to the simulation data for the insertion probability,
 \begin{equation}
  \phi = \frac{\rho}{k_p\tau},
 \end{equation}
 obtained from the measured steady--state densities for $0<k_p<600$. This procedure yields an optimal value $a\simeq 0.7795$.
\end{enumerate}

Figure~\ref{fig:yield} compares these three SPT--based models with the simulation data. All three provide a substantial improvement over the low--density virial expression~\eqref{eq:rho-low-density}. Model~(ii), which matches the exact $B_2$, is by construction accurate at small planting rates and remains surprisingly good up to $k_p\sim 10$, although it slightly overestimates the density at higher $k_p$. Model~(i) underestimates the density at large $k_p$. The one--parameter fit of model~(iii) yields an almost indistinguishable description of the data over the full range of planting rates considered. These results are consistent with the broader experience that SPT--inspired closures, suitably tuned, can provide accurate equations of state even for complex mixtures~\cite{santos2017equation,hansen2019scaling}.

\subsection{High planting--rate asymptotics}

The SPT--based description cannot be expected to be accurate in the dense limit, even for equilibrium hard disks, because it effectively assumes a maximal coverage of unity, whereas the true close--packing value is $\theta_{\rm cp} = \pi/(2\sqrt{3}) \simeq 0.907$~\cite{hansen2013theory}. For polydisperse hard disks, analytical results for random close packing are available~\cite{zaccone2025analytical}. In our system the steady--state density is bounded above by the optimal desynchronized value $\rho_{\max}=64/(27\sqrt{3})$~\cite{talbot2025optimising}, which plays a role analogous to a close--packing density.

To quantify the approach to this limit we fit the numerical steady--state densities to the empirical form
\begin{equation}
 \rho(k_p) = \rho_{\max} - \frac{A}{k_p^{\,b}},
 \label{eq:rho-asym}
\end{equation}
Performing a least-squares fit of $\log( \rho_{\max}-\rho)=\log A-b\log k_p$ to the simulation data yields $\log A=0.413$ and $b=0.333$, which is consistent with an exponent of $1/3$.

\begin{figure}[t]
 \centering
 \includegraphics[width=0.9\linewidth]{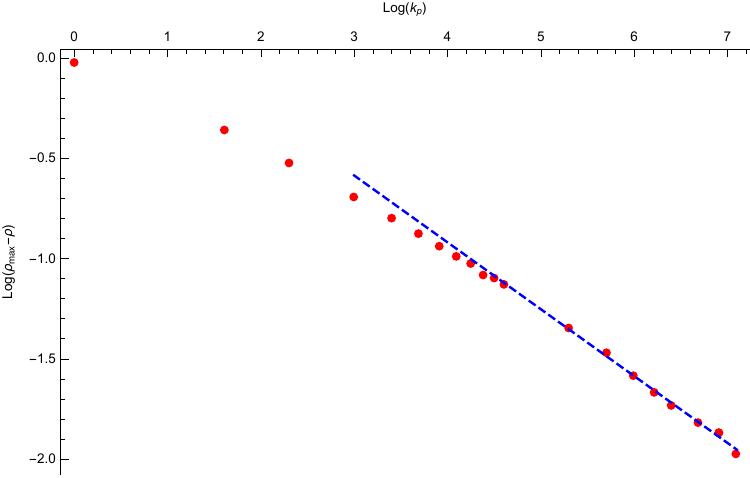}
 \caption{$\log( \rho_{\max}-\rho)$ versus $\log(k_p)$ in the steady--state density for $L/\sigma=20$ (points). Dashed line: linear fit of Eq.~\eqref{eq:rho-asym} with $A,b$ as free parameters and $\rho_{\max}$ fixed to $64/(27\sqrt{3})$ for $k_p\ge 100$. The best fit values are $\log(A)=0.413$ and $b=0.333$.}
 \label{fig:asymptotic}
\end{figure}

We do not presently have a theoretical explanation for the exponent $b\simeq 1/3$. It may be related to the statistics of voids and the rate at which the available area for new seeds shrinks at large planting rates, and it would be interesting to connect it to the scaling properties of packing--limited growth models~\cite{Dodds2002,dodds2003packing} and to rigorous results on random sequential packing and deposition~\cite{penrose2002limit}.

\section{Structure of the steady state}
\label{sec:structure}

\subsection{Radial distribution function}

The spatial structure of the field can be characterized by the radial distribution function $g(r)$, defined such that $2\pi r \rho g(r)\,{\rm d}r$ is the mean number of plants whose centers lie in an annulus of radius $r$ and width ${\rm d}r$ around a given plant. At low density $g(r)$ is related to the pair potential~\eqref{eq:pair-pot} via
\begin{equation}
 g(r) = \bigl\langle \exp[-\beta u(r,s_1,s_2)]\bigr\rangle_{s_1,s_2},
\end{equation}
where the average is taken over the steady--state radius distribution~$\psi(s)$. Using $\exp[-\beta u(r,s_1,s_2)]=\Theta[r-(\sigma-|s_1-s_2|)]$ and averaging over two independent radii uniformly distributed in $[0,\sigma/2]$ yields the exact low--density limit
\begin{equation}
 g_0(r) =
 \begin{cases}
  0, & 0 \le r < \sigma/2,\\[1ex]
  \bigl(1-2r/\sigma\bigr)^2, & \sigma/2 \le r \le \sigma,\\[1ex]
  1, & r > \sigma.
 \end{cases}
 \label{eq:g0}
\end{equation}

\begin{figure}[t]
 \centering
 \includegraphics[width=0.9\linewidth]{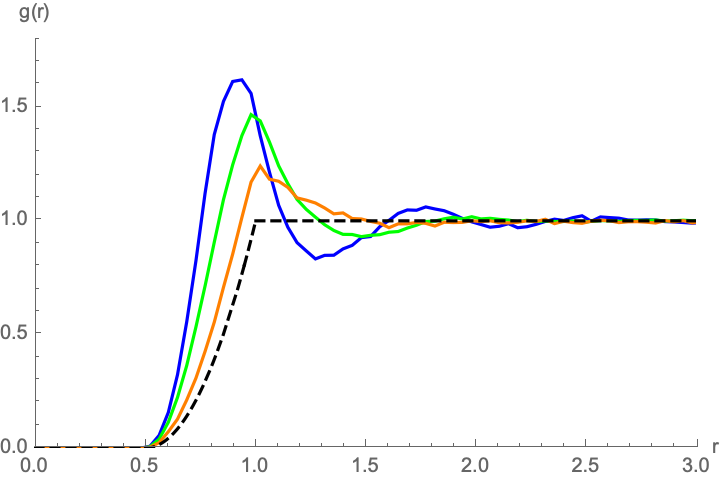}
 \caption{Radial distribution function $g(r)$ in the steady state for $k_p=1,10,100$. The dashed curve is the exact low--density limit~\eqref{eq:g0}.}
 \label{fig:gr}
\end{figure}

Figure~\ref{fig:gr} shows simulation results for $g(r)$ in the steady state for several planting rates, together with the exact low--density limit~\eqref{eq:g0}. As $k_p$ increases the first peak in $g(r)$ becomes more pronounced and shifts slightly, reflecting stronger local packing and the growing importance of correlations between plant ages. Nevertheless, the overall shape remains reminiscent of that of an equilibrium hard--disk fluid at comparable coverage~\cite{hansen2013theory}.

\subsection{Radius–resolved pair correlations}

To reveal correlations between plant sizes and spatial positions we introduce a radius--resolved pair correlation $g(z,r)$, where $z = |s_1-s_2|$ is the absolute difference in radii of two plants.

It is useful to distinguish two related but conceptually different descriptions. In the statistically stationary steady state of the random planting model, when configurations are sampled at arbitrary times, plant ages are uniformly distributed over $[0,\tau]$ for linear growth and the corresponding marginal radius distribution $\psi(s)$ is uniform on $[0,\sigma/2]$, independently of $k_p$. In that \emph{time-averaged} ensemble the baseline distribution of radius differences $p_0(z)$ is continuous. By contrast, in the dense, desynchronized configurations approached at large planting rates, the field becomes increasingly ordered and it is natural to consider a \emph{phase-conditioned} (stroboscopic) description in which configurations are sampled at a fixed phase of the growth--harvest cycle. In that setting the size distribution becomes sharply peaked and the corresponding $p_0(z)$ becomes discrete. The lattice reference discussed below and in Appendix~\ref{app:lattice-gzr} should be understood in this phase-conditioned sense.

We define the time-averaged $g(z,r)$ by
\begin{equation}
 {\rm d}N(z,r) = 2\pi r \rho\, p_0(z) g(z,r)\,{\rm d}r\,{\rm d}z,
 \label{eq:gzr-def}
\end{equation}
where ${\rm d}N(z,r)$ is the mean number of plants with radius difference in $[z,z+{\rm d}z]$ and separation in $[r,r+{\rm d}r]$, and $p_0(z)$ is the probability density of $z$ for two plants sampled independently from the steady--state radius distribution. For the uniform distribution~\eqref{eq:psi-linear} one finds
\begin{equation}
 p_0(z) = \frac{4}{\sigma}\left(1-\frac{2z}{\sigma}\right)
 = \frac{4}{\sigma}-\frac{8z}{\sigma^2},
 \qquad 0\le z \le \frac{\sigma}{2}.
\end{equation}
By construction $g(z,r)\to 1$ when correlations are absent.

\begin{figure}[t]
 \centering
 \includegraphics[width=0.9\linewidth]{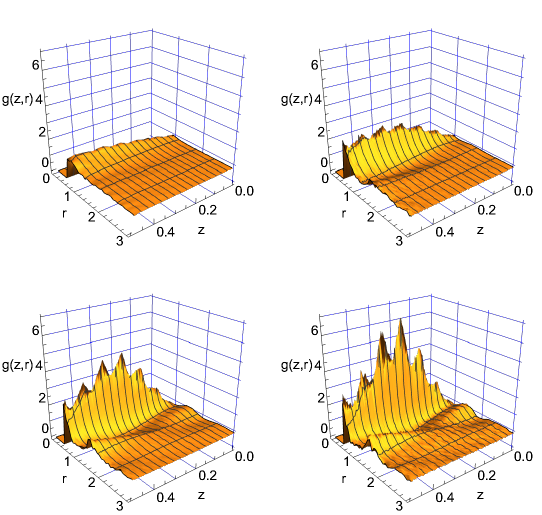}
 \caption{Radius–resolved pair correlation function $g(z,r)$ in the steady state for $L/\sigma=20$ and planting rates $k_p=1,10,100,500$ (left-to-right, top-to-bottom). A prominent ridge near $z\simeq\sigma/4$ at small~$r$ develops as $k_p$ increases, reflecting correlated parent–child pairs.}
 \label{fig:gzr}
\end{figure}

Figure~\ref{fig:gzr} shows $g(z,r)$ for $k_p=1,10,100,500$. At low planting rates the surface is relatively flat in~$z$, apart from the exclusion region at small~$r$, indicating weak correlations between the sizes of neighboring plants. As $k_p$ increases a ridge from $(z/\sigma,r/\sigma)\approx (0.5,0.5)$ to $(0,1)$ starts to develop.  At larger $k_p$ a pronounced maximum around $z\simeq \sigma/4$ appears on the ridge. This feature can be attributed to a seeding mechanism reminiscent of parent–child pairs. In a crowded field new seedlings are preferentially inserted just outside the exclusion halo of older plants. When a seedling is accepted in the vicinity of a parent with radius $s$, their age difference, and hence their radius difference, remain essentially constant as they grow. The probability that such a pair is observed at a given time is proportional to the product of the rate at which births occur near the parent’s boundary and the time during which the two plants coexist. This product is maximal for parents of intermediate size, leading to the observed peak near $z\simeq\sigma/4$. The spikes on the ridge appear to be artifacts due to the finite size of the system. Two additional ridges emerge as $k_p$ increases: one reaches a maximum at $z=0.5$, while the other is maximum at $z=0$.

The pronounced maximum on this ridge near $z\approx\sigma/4$
can be interpreted as a precursor of the mixed-size lattice structure shown in Fig.~\ref{fig:cones}. In the limit $k_p\rightarrow\infty$, the random planting dynamics approaches a desynchronized but highly ordered state in which plants are effectively organized into such mixed-size configurations. As emphasized by the spacetime representation in Fig.~\ref{fig:cones}, the two-dimensional lattice shown in the right panel corresponds to a fixed-time slice of a three-dimensional spacetime structure. As time advances, all plants grow, but their age differences—and hence the values of $z$
remain constant, as do the relative distances between neighboring pairs. Consequently, the ideal mixed-size lattice has a radius–resolved pair correlation 
$g(z,r)$ that is invariant over an entire growth cycle. The ridge and peak observed in the simulation data may therefore be viewed as the dynamically broadened signature of this underlying, time-independent geometric organization.

At larger separations $r$ the influence of direct parent–child relationships is diluted and $g(z,r)$ becomes almost independent of $z$, approaching the baseline $g(r)$. A detailed derivation of the lattice form of $g(z,r)$ and the corresponding
reference figure are given in Appendix~\ref{app:lattice-gzr}.

\section{Sigmoidal growth}
\label{sec:sigmoidal}

So far we have assumed linear growth in time. Real plants typically exhibit sigmoidal growth, for which a convenient parametrization is the logistic law
\begin{equation}
 r(t) = \frac{r_0}{1+\exp(-\alpha t + \beta)},
 \label{eq:logistic}
\end{equation}
where $r_0$ is the maximal radius, and $\alpha$ and $\beta$ control the growth rate and the inflection point. We again assume that plants are harvested when they reach a prescribed maturity radius $r_h = \sigma/2$ at time $t=\tau$, and that their initial radius at planting is $r_{\min}=c_p r_0$, with $0<c_p<c_h<1$ and $r_h=c_h r_0$.

The parameters $c_p$ and $c_h$ are related to $\alpha$ and $\beta$ by
\begin{equation}
 \beta = \ln\!\left(\frac{1}{c_p}-1\right), \qquad
 \alpha \tau = \beta - \ln\!\left(\frac{1}{c_h}-1\right).
\end{equation}
In the steady state the radius distribution $\psi(r)$ is proportional to the time that a plant spends near~$r$, i.e.,
\begin{equation}
 \psi(r) \propto \left(\frac{{\rm d}r}{{\rm d}t}\right)^{-1}.
\end{equation}
From Eq.~\eqref{eq:logistic} one obtains
\begin{equation}
 \frac{{\rm d}r}{{\rm d}t} = \alpha r_0 \left(\frac{r}{r_0}\right)\left(1-\frac{r}{r_0}\right),
\end{equation}
so that
\begin{equation}
 \psi(r) = \frac{a}{(r/r_0)\bigl(1-r/r_0\bigr)}, 
 \qquad c_p r_0 \le r \le c_h r_0,
\end{equation}
with a normalization constant~$a$ determined by
\begin{equation}
 \int_{c_p r_0}^{c_h r_0} \psi(r)\,{\rm d}r = 1.
\end{equation}
A short calculation yields
\begin{equation}
 a r_0 \ln\!\left[\frac{(1-c_p)c_h}{(1-c_h)c_p}\right] = 1.
\end{equation}

\begin{figure}[t]
 \centering
 \includegraphics[width=0.7\linewidth]{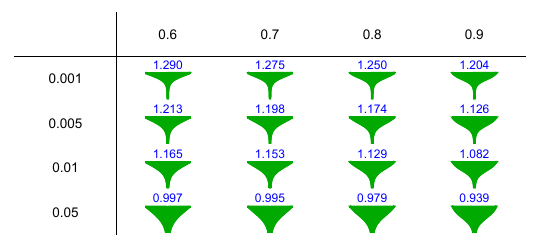}
 \caption{Second virial coefficient $B_2$ (in units of $\pi\sigma^2/4$) as a function of the logistic growth parameters $c_p$ (rows) and $c_h$ (columns), defined as the fractional radii at planting and harvest.}
 \label{fig:B2table}
\end{figure}

The mean second virial coefficient for a seedling interacting with a plant of radius~$r$ now depends on $c_p$ and $c_h$ through the radius distribution. Using Eq.~\eqref{eq:B2s} and averaging over $\psi(r)$ we obtain
\begin{equation}
 B_2 = \frac{\pi\sigma^2}{8 c_h^2}\, a r_0 
 \left[
  c_p - c_h + (4c_h-1)\ln\!\left(\frac{1-c_h}{1-c_p}\right)
  - 4 c_h^2 \ln\!\left(\frac{c_p(1-c_h)}{c_h(1-c_p)}\right)
 \right].
 \label{eq:B2-sigmoidal}
\end{equation}
Representative values of $B_2$ as a function of $c_p$ and $c_h$ are shown in Fig.~\ref{fig:B2table}. Inserting Eq.~\eqref{eq:B2-sigmoidal} into Eq.~\eqref{eq:rho-low-density} yields the corresponding low--density prediction for the steady--state density in the case of sigmoidal growth.

\section{Discussion and outlook}
\label{sec:discussion}

We have developed a statistical–mechanical description of a random planting model in which plants are represented by growing hard disks subject to a nonoverlap constraint over their entire lifetime. The central step is a mapping of the steady state onto a nonadditive polydisperse hard–disk fluid, which allows the steady-state planting probability to be interpreted as an insertion probability and related directly to an excess chemical potential.
At low density, this mapping leads naturally to a virial expansion for the insertion probability in terms of an effective second virial coefficient, which we computed explicitly for both linear and logistic growth laws. At moderate densities, we showed that a simple adaptation of scaled particle theory, involving a single effective diameter parameter, provides a quantitatively accurate prediction of the steady-state density over several decades of the planting rate. At very large planting rates, the density approaches the theoretically optimal value achieved by desynchronized regular planting~\cite{talbot2025optimising}, and the simulation data are consistent with an algebraic approach to this limit with an exponent close to $1/3$.

Beyond density and yield, we characterized the spatial organization of the field using both the conventional radial distribution function and a radius–resolved pair correlation $g(z,r)$. The latter reveals strong correlations between the sizes of nearby plants, including a pronounced enhancement of pairs whose radii differ by approximately one quarter of the maturity diameter. We interpret these correlations in terms of parent–child seeding events near the boundaries of exclusion halos. At high planting rates, the structure of $g(z,r)$ exhibits clear signatures of the same geometric constraints that underlie optimal desynchronized planting. In particular, the ridge and associated maxima observed in the simulations can be viewed as dynamically broadened precursors of the discrete mixed-size lattice that characterizes the ideal desynchronized configuration.

Several open questions remain. A theoretical explanation of the exponent governing the high-$k_p$ approach to the optimal density would be particularly desirable. Such an explanation may require a detailed analysis of the statistics of voids and available area in the effective packing-limited growth process generated by frequent planting attempts, and could benefit from connections to packing-limited growth models~\cite{Dodds2002,dodds2003packing} and to rigorous results on random sequential packing and deposition~\cite{penrose2002limit}. Extending the analytical treatment to truly polydisperse or multicomponent systems, in which different species have distinct growth laws and lifetimes, is another natural direction. Finally, it would be interesting to explore whether similar scaled-particle ideas can be applied to other nonequilibrium systems with dynamic exclusion constraints, such as models of tissue growth or deposition with restructuring~\cite{schaaf2000random,Kubala2022}.

\section{Data availability}
Programs used to simulate the model are openly available~\cite{GitHubLink}.

\section{Acknowledgment}

The author thanks David Colliaux and Pascal Viot for helpful comments on the manuscript.

\appendix
\section{Radius--resolved pair correlation for the ideal mixed--size lattice}
\label{app:lattice-gzr}

In this appendix we derive the radius--resolved pair correlation function $g(z,r)$
for the ideal mixed--size hexagonal configuration shown in Fig.~\ref{fig:cones},
which serves as a reference state for interpreting the simulation data in
Fig.~\ref{fig:gzr}. This configuration corresponds to the densest planting
arrangement obtained for a desynchronisation factor $\tau/2$ between successive
layers~\cite{talbot2025optimising}.

\subsection{Geometry of the mixed--size lattice}

In this idealized state, plants of radius $s_b=\sigma/2$ occupy a triangular
Bravais lattice with lattice parameter
\begin{equation}
 a=\frac{3\sqrt{3}}{4}\,\sigma,
 \label{eq:lattice-a}
\end{equation}
while plants of radius $s_g=\sigma/4$ occupy one of the interstitial sublattices.
There is one large and one small plant per primitive cell of area
$A_{\rm cell}=\sqrt{3}a^2/2$, so that the total number density is
\begin{equation}
 \rho=\frac{2}{A_{\rm cell}}=\frac{4}{\sqrt{3}\,a^2}.
\end{equation}

Only two values of the radius difference are possible,
\begin{equation}
 z = |s_1-s_2| =
 \begin{cases}
 0, & \text{same--size pairs},\\
 \sigma/4, & \text{mixed--size pairs},
 \end{cases}
\end{equation}
with equal probabilities $p_0(0)=p_0(\sigma/4)=1/2$ for two independently chosen
plants \emph{at fixed phase} in this idealized, desynchronized state..

\subsection{Definition of the lattice $g(z,r)$}

For a perfect infinite lattice, the radius--resolved pair correlation function
is a sum of delta--function contributions,
\begin{equation}
 g(z,r)
 = \sum_n A_n^{(0)}\,\delta(z)\,\delta(r-r_n^{(0)})
 + \sum_m A_m^{(1)}\,\delta\!\left(z-\frac{\sigma}{4}\right)\delta(r-r_m^{(1)}),
 \label{eq:gzr-lattice}
\end{equation}
where $r_n^{(0)}$ and $r_m^{(1)}$ denote the shell distances for same--size and
mixed--size pairs, respectively. The amplitudes are fixed by the definition
\begin{equation}
 \mathrm{d}N(z,r)
 = 2\pi r\,\rho\,p_0(z)\,g(z,r)\,\mathrm{d}r\,\mathrm{d}z,
 \label{eq:gzr-def-app}
\end{equation}
so that a shell at distance $r$ with multiplicity $n$ contributes an amplitude
\begin{equation}
 A = \frac{n}{2\pi r\,\rho\,p_0(z)}.
 \label{eq:lattice-amplitude}
\end{equation}

\subsection{First lattice shells}

For same--size pairs $(z=0)$, each sublattice is itself a triangular lattice.
The first shell distances are
\begin{equation}
 r_1^{(0)} = a, \qquad
 r_2^{(0)} = \sqrt{3}\,a, \qquad
 r_3^{(0)} = 2a, \ldots,
\end{equation}
each with multiplicity $n=6$.

For mixed--size pairs $(z=\sigma/4)$, the shell distances are determined by the
relative displacement between the two sublattices. The nearest mixed shell
occurs at the minimal allowed separation imposed by the exclusion rule,
\begin{equation}
 r_1^{(1)} = d_{\min}
 = \sigma-|s_b-s_g| = \frac{3\sigma}{4},
\end{equation}
with $s_b=\sigma/4,s_g=\sigma/2$ and multiplicity $n=3$. Higher mixed shells appear at larger, discrete values
of $r$, with multiplicities determined by the lattice geometry.

Using Eqs.~\eqref{eq:lattice-a} and \eqref{eq:lattice-amplitude}, the first few
delta--function contributions to $g(z,r)$ can be written explicitly for the first two shells of each type of pair ($z=0,\sigma/4$):
\begin{align}
g(\sigma/4,r)=\frac{27\sqrt{3}}{16\pi}\sigma\delta(r-\frac{3}{4}\sigma)+\frac{27\sqrt{3}}{32\pi}\sigma\delta(r-\frac{3}{2}\sigma)\\
g(0,r)=\frac{27}{8\pi}\sigma\delta(r-\frac{3\sqrt{3}}{4}\sigma)+\frac{9\sqrt{3}}{8\pi}\sigma\delta(r-\frac{9}{4}\sigma)
\end{align}

\begin{figure}[t]
 \centering
 \includegraphics[width=0.8\linewidth]{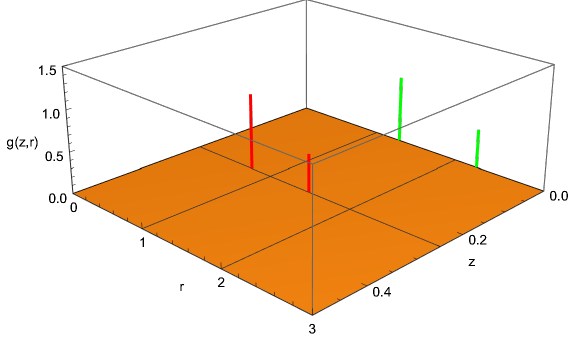}
 \caption{Radius--resolved pair correlation function $g(z,r)$ for the ideal
 mixed--size hexagonal lattice corresponding to Fig.~\ref{fig:cones}. Shown are
 the first four delta--function peaks for same--size $(z=0)$ and mixed--size
 $(z=\sigma/4)$ pairs. These peaks provide a geometric reference for interpreting
 the ridge structure observed in Fig.~\ref{fig:gzr} for the random planting model
 at large planting rates.}
 \label{fig:gzr_lattice}
\end{figure}

These are shown schematically in Fig.~\ref{fig:gzr_lattice}.

\subsection{Relation to the random planting model}

In the random planting model the steady state is disordered and the radius
distribution is continuous, so the delta peaks of
Eq.~\eqref{eq:gzr-lattice} are broadened into extended features.
Nevertheless, the hard--core condition implies that for a given radius difference
$z$ the minimal admissible separation is $r=\sigma-z$. As the planting rate
$k_p$ increases, accepted seedlings accumulate near this boundary, producing
the ridge observed in Fig.~\ref{fig:gzr}. The pronounced maximum near
$z\simeq\sigma/4$ corresponds to the first mixed--size lattice shell at
$r=3\sigma/4$, while the extension of the ridge toward larger $r$ reflects
higher--order lattice shells.

As emphasized by the spacetime representation in Fig.~\ref{fig:cones}, the
two--dimensional lattice shown in that figure is a fixed--time slice of a
three--dimensional spacetime structure. As time increases, all plants grow but
their age differences—and hence the values of $z$—remain constant, as do the
distances between neighboring pairs. Consequently, the ideal lattice
$g(z,r)$ is invariant over an entire growth cycle. The simulation data therefore
represent a dynamically broadened precursor of this time--independent mixed--size
lattice structure, approached asymptotically in the limit $k_p\to\infty$.

\end{document}